# An Alternative Model of Quark Confinement

## Hadron Fields

**Dr Charles Francis**

## Abstract

In [1] we constructed a form of field theory in which Feynman diagrams describe real particle interactions, not virtual ones. In this paper we outline a theory of discrete interactions based on hadron field operators which bind quark in co-ordinate space, but not momentum space, and offer explanations for the strong force, jets and Zweig's rule. As yet no calculations have been carried out on the predictions of the theory, but we indicate possibilities for testable predictions.

28/9/99

Dr Charles Francis
Lluest, Neuaddlwyd
Lampeter
Ceredigion
SA48 7RG

# An Alternative Model of Quark Confinement

## 1    Background

We have previously given empirical [1] and metaphysical [2] approaches to discrete quantum electrodynamics. As in the standard model, photons transmit the electromagnetic force and massive vector bosons transmit the weak interaction as in the Weinberg-Salaam-Glashow theory [3], but the Lagrangian describes interactions between real point-like particles, not virtual ones. In the model elementary bosons (the photon, W and Z) are destroyed in one cycle of their own time-line, and other stable (or mainly stable) elementary particles are Dirac particles. The assumption is that quarks are Dirac particles and that all of their properties are described by the Dirac equation together with their interactions with bosons.

Interactions between real particles are described by constructing quantum field operators on the vector space of measurable states. In the present paper we consider hadronic field operators which bind quarks in co-ordinate space, but not in momentum space. The theory offers explanations for the strong force, jets and Zweig's rule, while remaining consistent with successful predictions of the behaviour of hadrons in other theories. For example, because of symmetries and the constraints of relativity and quantum mechanics on the S-matrix [4] it may be that cross-sections predicted by this model are the same as the successful predictions of QCD [5] and string theory [6][7], but the metaphysic suggested here is considerably simpler does not leave the unanswered questions of those models.

We have not carried out quantitive calculations. This paper is an outline study of the main features of the model and a qualitive discussion of possible interactions which show properties of hadrons. Initially we consider a purely hadronic interaction, generating a Yukawa type potential [8] but it will be seen when the electromagnetic interaction is considered that it also contributes to quark confinement and also to the strong force, so the coupling constant of the hadronic interaction is much less than the Yukawa value, and it is possible that perturbation theory can be used.

There appears to be scope for testing the model in the analysis of jets and in the predictions of the gyromagnetic moment of the proton and the neutron. Jets are are predicted to be formed out the the creation of hadron-antihadron pairs, not out of individual quarks or gluons. It is thought that jets will show some characteristic from the original particle production, and that such characteristics can in principle be analysed. On the grounds of a general similarity between this model and QCD we anticipate a value for the gyromagnetic moment of the proton and the neutron which is near to, but not the same as, the prediction of QCD. But, in common with QCD, to test it a very substantial calculation must be carried out without error. The calculation has not been attempted at the present time.

We will use the following notations defined in [1] and [2]. $S$ is the set of spin indices. $\mathscr{F}$ is the set of physically realisable states. The coordinate system is $N = (-v, v] \otimes (-v, v] \otimes (-v, v] \subset \mathbb{N}^3$ for some $v \in \mathbb{N}$, where $(-v, v] = \{x \in \mathbb{N} \mid -v < x \leq v\}$. A field operator is a mapping $\mathbb{R}^4 \otimes S \to \mathscr{F}$, where the elements of $\mathscr{F}$ are regarded as operators. The field operator for the creation of a particle in interaction is

1.1 $\qquad \forall (x, \alpha) = (x_0, \boldsymbol{x}, \alpha) \in \mathbb{R}^4 \otimes S$ . $\underline{|x, \alpha\rangle} {:} \mathscr{F} \to \mathscr{F}$

The field operators for the of creation of the antiparticle is $\overline{|x, \alpha\rangle} {:} \mathscr{F} \to \mathscr{F}$ . The field operators for the annihilation of particles in interaction are $\underline{\langle x, \alpha|} {:} \mathscr{F} \to \mathscr{F}$ and $\langle \overline{x, \alpha}| {:} \mathscr{F} \to \mathscr{F}$ and are the hermitian conjugates of the creation field operators.

## 2  Hadron Fields

It is postulated that all interactions can be represented as a number of particles annihilated or created at a point. Dirac particles are fermions, obeying the Pauli exclusion principle so an interaction cannot create an even number of the same Dirac particle. But there is nothing to stop the creation of a particle-antiparticle pair, or the creation of a triplet of three Dirac particles (provided that there is no two particle subsystem).

**Definition:** The baryon field operator creating three quarks, indexed 1,2,3, at $x \in \mathbb{R}^4$ is

2.1
$$| \, x \, , \underline{\alpha}_1, \underline{\alpha}_2, \underline{\alpha}_3 \rangle \equiv \underline{| \, x \, , \alpha_1, \alpha_2, \alpha_3 \rangle} \equiv \underline{| \, x \, , \alpha_1 \rangle} \underline{| \, x \, , \alpha_2 \rangle} \underline{| \, x \, , \alpha_3 \rangle}$$

**Definition:** The baryon field operator creating three antiquarks, indexed 1,2,3, at $x \in \mathbb{R}^4$ is

2.2
$$| \, x \, , \overline{\alpha}_1, \overline{\alpha}_2, \overline{\alpha}_3 \rangle \equiv \overline{| \, x \, , \alpha_1, \alpha_2, \alpha_3 \rangle} \equiv \overline{| \, x \, , \alpha_1 \rangle} \, \overline{| \, x \, , \alpha_2 \rangle} \, \overline{| \, x \, , \alpha_3 \rangle}$$

**Definition:** The meson field operator creating a quark, 1 and antiquark, 2 is

2.3
$$| \, x \, , \underline{\alpha}_1, \overline{\alpha}_2 \rangle \equiv \underline{| \, x \, , \alpha_1, \alpha_2 \rangle} \equiv \underline{| \, x \, , \alpha_1 \rangle} \overline{| \, x \, , \alpha_2 \rangle}$$

Interacting field operators make it explicit that in interaction the annihilation of a particle is equivalent to the creation of the corresponding antiparticle. They are regularised so that a quark cannot be created and annihilated in the same interaction.

**Definition:** For a baryon the interacting field operator is

2.4
$$\phi_{\alpha_1 \alpha_2 \alpha_3}(x) \; = \; | \, x \, , \overline{\alpha}_1, \overline{\alpha}_2, \overline{\alpha}_3 \rangle + \langle \, x \, , \underline{\alpha}_1, \underline{\alpha}_2, \underline{\alpha}_3 |$$

**Corollary:** The hermitian conjugate of the interacting baryon field is the antibaryon field

2.5
$$\phi^\dagger{}_{\alpha_1 \alpha_2 \alpha_3}(x) \; = \; \phi_{\overline{\alpha}_1 \overline{\alpha}_2 \overline{\alpha}_3}(x)$$

**Proof:** immediate from 2.4.

**Definition:** For a meson the interacting field operator is

2.6
$$\phi_{\alpha_1 \overline{\alpha}_2}(x) \; = \; | \, x \, , \overline{\alpha}_1, \underline{\alpha}_2 \rangle + \langle \, x \, , \underline{\alpha}_1, \overline{\alpha}_2 |$$

**Corollary:** The hermitian conjugate of the interacting meson field is also a meson field

2.7
$$\phi^\dagger{}_{\alpha_1 \overline{\alpha}_2}(x) \; = \; \phi_{\overline{\alpha}_1 \alpha_2}(x) \; = \; \phi_{\alpha_2 \overline{\alpha}_1}(x)$$

**Proof:** immediate from 2.6  and 2.3.

It is possible to postulate purely hadronic interactions from these field operators, corresponding to the Yukawa potential [8] in which a baryon absorbs or emits a meson (figure 2.9). Then the general form of the interaction operator is

2.8
$$\sum_{x \, \in \, \mathbb{N}} \phi_{\overline{\alpha}_1 \overline{\alpha}_2 \overline{\alpha}_3}(x) \phi_{\beta_3 \overline{\alpha}_4}(x) \phi_{\beta_1 \beta_2 \beta_4}(x) \Gamma_{\alpha_1 \alpha_2 \alpha_3 \alpha_4 \beta_1 \beta_2 \beta_3 \beta_4}$$

plus terms for the interaction with quarks in the first and second index positions. The summation convention is assumed for spin indices, and $\Gamma$ is some combination of Dirac $\gamma$ matrices. All possible quark combinations are included, subject to the condition that quark number is conserved in the interaction for each flavour of quark.

**Figure 2.9:** *Feynman diagram for a possible interaction in which a baryon absorbs or emits a meson. The (optional) heavy joining line indicates the particles meet at a point.*

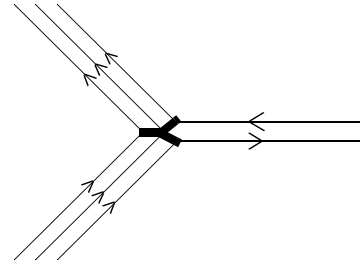

The corresponding meson interaction (figure 2.11) is

2.10 $$\sum_{x \in N} \phi_{\bar{\alpha}_1\alpha_2}(x)\phi_{\bar{\beta}_2\alpha_3}(x)\phi_{\beta_1\bar{\beta}_3}(x)\Gamma_{\alpha_1\alpha_2\alpha_3\beta_1\beta_2\beta_3}$$

**Figure 2.11:** *Feynman diagram for a possible interaction in which a meson absorbs or emits a meson.*

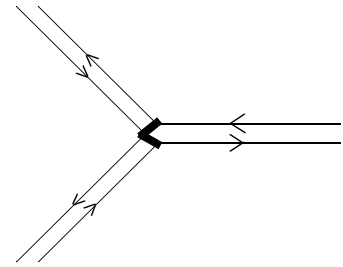

# 3   The Electromagnetic Interaction

The treatment of the electromagnetic interaction follows that for leptons, but there are now three Dirac adjoints for baryons, and two for mesons.

**Definition:** The Dirac adjoints of the baryon annihilation operator $\langle x, \underline{\alpha}_1, \underline{\alpha}_2, \underline{\alpha}_3 |$ are

3.1 $\qquad | x, \hat{\underline{\alpha}}_1, \underline{\alpha}_2, \underline{\alpha}_3 \rangle = | x, \underline{\mu}, \underline{\alpha}_2, \underline{\alpha}_3 \rangle \gamma^0_{\mu\alpha_1}$ $\qquad$ (summation convention is used for $\mu$)

3.2 $\qquad | x, \underline{\alpha}_1, \hat{\underline{\alpha}}_2, \underline{\alpha}_3 \rangle = | x, \underline{\alpha}_1, \underline{\mu}, \underline{\alpha}_3 \rangle \gamma^0_{\mu\alpha_2}$

3.3 $\qquad | x, \underline{\alpha}_1, \underline{\alpha}_2, \hat{\underline{\alpha}}_3 \rangle = | x, \underline{\alpha}_1, \underline{\alpha}_2, \underline{\mu} \rangle \gamma^0_{\mu\alpha_3}$

**Definition:** The Dirac adjoints of the meson annihilation operator $\langle x, \underline{\alpha}_1, \bar{\alpha}_2 |$ are

3.4 $\qquad | x, \underline{\alpha}_1, \hat{\bar{\alpha}}_2 \rangle = | x, \underline{\alpha}_1, \bar{\mu} \rangle \gamma^0_{\mu\alpha_2}$

3.5 $\qquad | x, \hat{\underline{\alpha}}_1, \bar{\alpha}_2 \rangle = | x, \underline{\mu}, \bar{\alpha}_2 \rangle \gamma^0_{\mu\alpha_1}$

**Definition:** The Dirac adjoints of the baryon field $\phi_{\alpha_1\alpha_2\alpha_3}(x)$, 2.4, are

3.6 $\qquad \phi_{\hat{\bar{\alpha}}_1\alpha_2\bar{\alpha}_3}(x) = \phi^\dagger_{\mu\alpha_2\alpha_3}(x)\gamma^0_{\mu\alpha_1} = | x, \hat{\underline{\alpha}}_1, \underline{\alpha}_2, \underline{\alpha}_3 \rangle + \langle x, \hat{\bar{\alpha}}_1, \bar{\alpha}_2, \bar{\alpha}_3 |$

and similarly for the other two quarks $\phi_{\bar{\alpha}_1\hat{\alpha}_2\bar{\alpha}_3}(x)$ and $\phi_{\bar{\alpha}_1\hat{\alpha}_2\bar{\alpha}_3}(x)$.

**Definition:** The Dirac adjoints of the meson field $\phi_{\alpha_1\bar{\alpha}_2}(x)$, 2.6, are

3.7 $\qquad \phi_{\hat{\bar{\alpha}}_1\alpha_2}(x) = (| x, \underline{\mu}, \bar{\alpha}_2 \rangle + \langle x, \bar{\mu}, \underline{\alpha}_2 |)\gamma^0_{\mu\alpha_1}$

3.8 $\qquad \phi_{\bar{\alpha}_1\hat{\alpha}_2}(x) = (| x, \underline{\alpha}_1, \bar{\mu} \rangle + \langle x, \bar{\alpha}_1, \underline{\mu} |)\gamma^0_{\mu\alpha_1}$

The coupling of photon to quark in an electromagnetic interaction follows the pattern of the coupling of photon to lepton in QED. The photon only couples to one quark at a time, but all three quarks of a baryon, or both quarks of a meson are part of the field. Let the quarks indexed 1,2,3 have charges $e_1$, $e_2$, $e_3$. Then the electromagnetic interaction for a baryon (figure 3.10) is

3.9
$$\sum_{x \in N} :e_1 \phi_{\hat{\overline{\alpha}}_1 \overline{\alpha}_2 \overline{\alpha}_3}(x) \gamma_{\alpha_1 \alpha'_1} \cdot A(x) \phi_{\alpha'_1 \alpha_2 \alpha_3}(x)$$
$$+ e_2 \phi_{\overline{\alpha}_1 \hat{\overline{\alpha}}_2 \overline{\alpha}_3}(x) \gamma_{\alpha_2 \alpha'_2} \cdot A(x) \phi_{\alpha_1 \alpha'_2 \alpha_3}(x)$$
$$+ e_3 \phi_{\overline{\alpha}_1 \overline{\alpha}_2 \hat{\overline{\alpha}}_3}(x) \gamma_{\alpha_3 \alpha'_3} \cdot A(x) \phi_{\alpha_1 \alpha_2 \alpha'_3}(x):$$

where the colons reorder the expression by placing all creation operators to the left of all annihilation operators, to ensure that false values are not generated corresponding to the annihilation of particles in the interaction in which they are created.

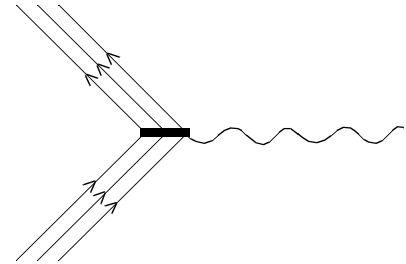

**Figure 3.10:** *Feynman diagram for an interaction in which a baryon absorbs or emits a photon. The photon couples to one quark, but all three quarks must be present.*

Similarly for a meson the electromagnetic interaction (figure 3.12) is

3.11
$$\sum_{x \in N} :e_1 \phi_{\hat{\overline{\alpha}}_1 \alpha_2}(x) \gamma_{\alpha_1 \alpha'_1} \cdot A(x) \phi_{\alpha'_1 \overline{\alpha}_2}(x)$$
$$- e_2 \phi_{\overline{\alpha}_1 \hat{\alpha}_2}(x) \gamma_{\alpha_2 \alpha'_2} \cdot A(x) \phi_{\alpha_1 \overline{\alpha}'_2}(x):$$

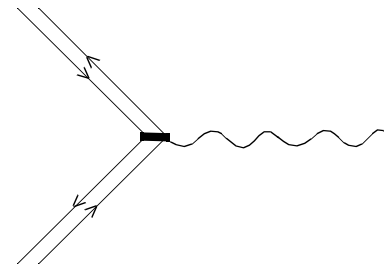

**Figure 3.12:** *Feynman diagram for an interaction in which a meson absorbs or emits a photon. The photon couples to one quark, but both quarks must be present.*

**Theorem:** In the electromagnetic interaction, in the case when the hadron is preserved (i.e. ignoring pair creation and pair annihilation) the photon interacts with one quark, and leaves spin and momentum of the other quark(s) unchanged.

**Proof:** It is sufficient to demonstrate the theorem for the first term of 3.9. The proof for the other terms of 3.9 and 3.11 is identical. The first term of 3.9 is

3.13
$$\sum_{x \in N} e_1 \phi_{\hat{\overline{\alpha}}_1 \overline{\alpha}_2 \overline{\alpha}_3}(x) \gamma_{\alpha_1 \alpha'_1} \cdot A(x) \phi_{\alpha'_1 \alpha_2 \alpha_3}(x)$$

Then using 3.6 and 2.4, 3.13 can be written

3.14
$$\sum_{x \,\in\, \mathrm{N}} : e_1 (|\,x\,,\hat{\underline{\alpha}}_1,\underline{\alpha}_2,\underline{\alpha}_3\rangle + \langle\,x\,,\hat{\bar{\alpha}}_1,\bar{\alpha}_2,\bar{\alpha}_3|)\gamma_{\alpha_1\alpha'_1}\cdot A(x)$$
$$(|\,x\,,\bar{\alpha}'_1,\bar{\alpha}_2,\bar{\alpha}_3\rangle + \langle\,x\,,\underline{\alpha}'_1,\underline{\alpha}_2,\underline{\alpha}_3|):$$

The term affecting a particle is

3.15
$$\sum_{x \,\in\, \mathrm{N}} e_1 |\,x\,,\hat{\underline{\alpha}}_1,\underline{\alpha}_2,\underline{\alpha}_3\rangle \gamma_{\alpha_1\alpha'_1}\cdot A(x)\langle\,x\,,\underline{\alpha}'_1,\underline{\alpha}_2,\underline{\alpha}_3|$$

which is, by 2.1, in matrix notation ([1] section 6, *Multiparticle States*).

3.16
$$\begin{bmatrix} \displaystyle\sum_{x \,\in\, \mathrm{N}} e_1 |\,x\,,\hat{\underline{\alpha}}_1\rangle \gamma_{\alpha_1\alpha'_1}\cdot A(x)\langle\,x\,,\underline{\alpha}'_1| \\[2mm] \displaystyle\sum_{x \,\in\, \mathrm{N}} |\,x\,,\underline{\alpha}_2\rangle\langle\,x\,,\underline{\alpha}_2| \\[2mm] \displaystyle\sum_{x \,\in\, \mathrm{N}} |\,x\,,\underline{\alpha}_3\rangle\langle\,x\,,\underline{\alpha}_3| \end{bmatrix}$$

In 3.16 it is seen that the first quark participates in an interaction with an identical form to the electromagnetic interaction for leptons, while the other two particles are unaffected, by the resolution of unity. An identical argument applies to the antiparticle term in 3.14, and to each of the terms in 3.9 and 3.11.

## 4    Consequences

### 4.1   The Strong Field

Just as the classical electromagnetic field consists of a cloud of "virtual" photons surrounding a charged particles, in the present model the strong field consists of "virtual" hadrons (principally mesons) surrounding a hadron. Virtual particles are perhaps more accurately thought of as real. The absence of knowledge about any individual virtual particles implies that they have a probabilistic effect on the behaviour of matter, which can be described in a field theoretic approach. The distinction between this and QCD is that here the strong field consists solely of quarks and anti-quarks created in interactions of the form of 2.8, 2.10, 3.9, and 3.11. We are not aware of any direct experimental test which can be carried out for the absence of gluons.

### 4.2   Isospin and SU(N)

The properties of SU(N) follow directly and immediately from the definition of multiparticle space given in [1], and the interactions 2.8 and 2.10 under the constraint on that possible quark combinations are included subject to the condition that quark number is conserved in the interaction for each flavour of quark. SU(N) is equivalent to the statement that in the absence of knowledge of the flavour of an individual quark, it is labelled by a ket formed from a linear combination of the possible flavours.

### 4.3   Zweig's Rule

It has been observed in experiment that, when a meson consists of a quark and its corresponding anti-quark, the two do not immediately self-destruct. This follows directly from the interaction 2.10. As a result the mesons consisting of $s\overline{s}$, $c\overline{c}$, and $b\overline{b}$ combinations are relatively stable. Conservation of energy prevents their decay into other particles in which the quark and antiquark survive separately

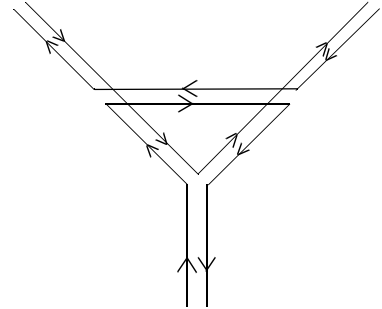

**Figure 4.4:** *Zweig's rule is not absolute, because the quark-anti-quark pair can annihilate in a second order process. This requires greater energy than was available in the original meson, but takes place because conservation of energy is temporarily suspended.*

### 4.5   Superstrong Force

Quarks are confined in hadrons because there is only one space co-ordinate in the interaction density for all the quarks in an interaction 2.8, 2.10, 3.9 and 3.11. It is convenient to think that they share the same wave function in co-ordinate space; if one quark is localised, for example by an electromagnetic interaction with other matter, then the other quarks are confined at the same point when the interaction takes place. These interactions are taking place all the time so quarks are confined within baryons and mesons. Free quarks, if they exist, cannot participate in electromagnetic interactions, and therefore cannot be directly detected, but if they do exist we might expect certain interactions to take place more easily, and this may be testible.

### 4.6   Quasi-Free Quarks

The interaction generates a momentum conserving delta function for each individual quark. Thus for hadrons in which the quarks are in eigenstates of momentum, the interaction leaves the momenta and the spin states of two quarks in a baryon unchanged, as though only one of the three quarks participates in the interaction. Thus quarks are `quasi free' – they are confined in co-ordinate space, and yet have values of momentum independent of each other.

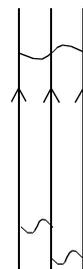

**Figure 4.7:** *Photons passing between the quarks in a baryon transfer momentum from one quark to another.*

The self interaction in which a photon is exchanged between two of the quarks in the baryon ensures that momentum can be transferred from one quark to another (figure 4.7). Quasi free quarks can also be understood by application of the uncertainty principle. If the quarks in a hadron are confined in coordinate space, their momenta are indeterminate, and hence independent. The prediction that quarks are quasi-free was originally observed experimentally, but in the absence of any other satisfactory model of quark confinement, we believe that it strongly supports the current model.

### 4.8 Quark Sharing

Only quark triplets and quark-antiquark pairs can interact, so only these combinations can be observed. But this does not resolve whether quarks are independent particles or whether they are parts of more complex particles with an indivisible structure. The resolution of this question depends on whether, once a quark triplet has been created, the same three quarks must come together to be annihilated in the next interaction, or whether any three quarks would do. We will now examine the qualitive properties of hadrons on the assumption that interactions can consist of any three quarks.

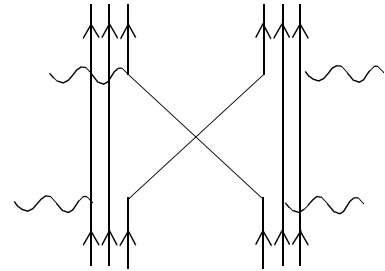

**Figure 4.9:** *If quarks are independent particles then it need not be the same quark triplets which participate in all interactions in the nucleus.*

The result of allowing any triplet of free quarks to combine in an interaction for a baryon is that quarks will be shared between baryons in the nucleus of the atom. Effectively this results in the transmission of mesons between hadrons, and conveys a Yukawa-type potential by means of electromagnetic interactions, considerably amplifying the strong force, or rather reducing the value of the coupling constant necessary to maintain it. Because the quarks in a hadron are closely bound it is expected that large numbers of photons are transmitted between them, and it is not immediately obvious how to calculate cross-sections or to predict magnitude of the strong binding force. In addition to electromagnetic interactions, the calculation of the strong force may involve purely hadronic interactions, as well as "identity interactions"

$$4.10 \qquad \sum_{x \,\in\, \mathbb{N}} \phi_{\overline{\alpha}_1 \overline{\alpha}_2 \overline{\alpha}_3}(x) \phi_{\alpha_1 \alpha_2 \alpha_3}(x) \ \text{ and } \ \sum_{x \,\in\, \mathbb{N}} \phi_{\overline{\alpha}_1 \alpha_2}(x) \phi_{\alpha_1 \overline{\alpha}_2}(x)$$

which enable quark sharing without the need for a photon. As with the second an third entries of 3.16, identity interactions leave the momenta and spin states of the individual quarks unchanged, while still requiring them to be confined in space.

### 4.11 Jets in Hadron Scattering

If hadrons are fired at each other at very high energies, they may exchange a meson (figure 4.12). The quarks within the resulting pair of hadrons have excessive energy differences and do not have a stable shared waved function. Thus the resulting hadrons rapidly decay. Because each of the two hadrons has an overall momentum along the axis the original collision, they decay into jets of hadrons moving on that axis. The energy and spread of the jets should be calculable from a knowledge of the (stable) state of quarks in the initial hadrons.

**Figure 4.12:** *Meson exchange by the exchange of two quarks..*

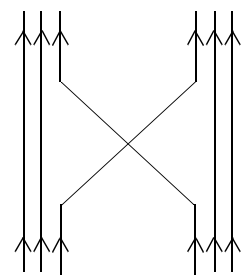

### 4.13 Jets in Electron Positron Scattering

A beam of positrons is fired at a beam of electrons at very high energy. A positron combines with an electron to create a single photon. The photon has zero momentum and seriously violates conservation of energy, so it must decay rapidly into other particles.

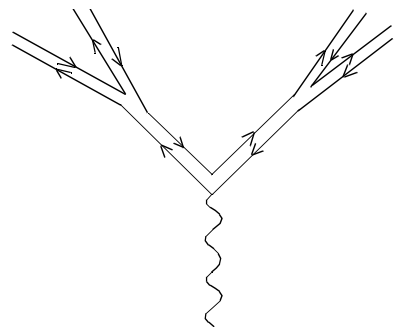

**Figure 4.14:** *An energy conservation violating photon creates a hadron antihadron pair. The hadrons have equal and opposite total momenta, and high internal energy, so they rapidly decay into jets of particles in opposite directions.*

This explanation of jets appears to us more successful than the usual interpretation in quantum chromodynamics, where it is suggested that jets arise from a single quark, and which fails to satisfactorily resolve the problem that if the jet arose from a single quark there should be a loose quark left over after the formation of hadrons. We expect jets not from single quarks, but as a result of any interaction which seriously violates energy conservation, as the interaction initially creates particles with high momenta to restore energy conservation as quickly as possible.

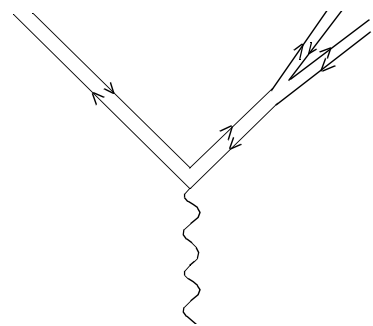

**Figure 4.15:** *According to the model, at high energies greater internal energy is dissipated in the first decay after the initial pair creation. I.e. this interaction does more to restore conservation of energy than subsequent ones. As energies increase, the residual inbalance in energy conservation first leads to a thickening of one of the jets, and then to a separate jet.*

If this model of jets is correct, then we expect to categorise patterns of jets according to the initial pair creation. A frequent decay would derive from a mesons consisting of two quarks of approximately equal mass. In this case two of the jets would derive from one meson and be approximately equal, and the other would tend to be of nearly double the energy. A less frequent occurence would be jets arising from the creation of an initial baryon-antibaryon pair, where the third jet will corresponding to a meson splitting of a baryon, and will generate a weaker jet. At any given energy, the frequency of each pattern of jets should correspond to the predicted frequency of each hadron pair creation, and can be calculated by standard methods.

### 4.16 Perturbation Theory.

In [1] we calculated Feynman rules and showed that the current discrete model of qed necessitates that we subtract a term the propagator which recognises that a particle cannot be annihilated at the instant of its creation. It was shown that when this term is subtracted finite results are a obtained in agreement with the standard predictions of qed after renormalisation. It is necessary to renormalise the electron mass to take into account the self interaction, but no infinite quantities are involved. The perturbation theoretic approach based on Feynman diagrams used in this paper gives qualitive agreement with experimental results. The mechanisms proposed here for quark binding and quark sharing predict a coupling constant with a value much less than that anticipated by the Yukawa potential based on simple meson exchange e.g. [9], so a perturbation theory approach to calculation may be technically feasible. But, although the metaphysic is much simpler than that of qcd or string theory there are still large numbers of diagrams and quantitive predictions are not so simple. In other respects the model shares predictions with standard theories based on general considerations such as analyticy, relativistic invariance, and symmetries, as well as the successful phenomenology other parton models [5].

# An Alternative Model of Quark Confinement

## Hadron Fields

## Dr Charles Francis

## Abstract

In [1] we constructed a form of field theory in which Feynman diagrams describe real particle interactions, not virtual ones. In this paper we outline a theory of discrete interactions based on hadron field operators which bind quark in co-ordinate space, but not momentum space, and offer explanations for the strong force, jets and Zweig's rule. As yet no calculations have been carried out on the predictions of the theory, but we indicate possibilities for testable predictions.

Dr Charles Francis
Lluest, Neuaddlwyd
Lampeter
Ceredigion
SA48 7RG

# An Alternative Model of Quark Confinement

## 1    Background

We have previously given empirical [1] and metaphysical [2] approaches to discrete quantum electrodynamics. As in the standard model, photons transmit the electromagnetic force and massive vector bosons transmit the weak interaction as in the Weinberg-Salaam-Glashow theory [3], but the Lagrangian describes interactions between real point-like particles, not virtual ones. In the model elementary bosons (the photon, W and Z) are destroyed in one cycle of their own time-line, and other stable (or mainly stable) elementary particles are Dirac particles. The assumption is that quarks are Dirac particles and that all of their properties are described by the Dirac equation together with their interactions with bosons.

Interactions between real particles are described by constructing quantum field operators on the vector space of measurable states. In the present paper we consider hadronic field operators which bind quarks in co-ordinate space, but not in momentum space. The theory offers explanations for the strong force, jets and Zweig's rule, while remaining consistent with successful predictions of the behaviour of hadrons in other theories. For example, because of symmetries and the constraints of relativity and quantum mechanics on the S-matrix [4] it may be that cross-sections predicted by this model are the same as the successful predictions of QCD [5] and string theory [6][7], but the metaphysic suggested here is considerably simpler does not leave the unanswered questions of those models.

We have not carried out quantitive calculations. This paper is an outline study of the main features of the model and a qualitive discussion of possible interactions which show properties of hadrons. Initially we consider a purely hadronic interaction, generating a Yukawa type potential [8] but it will be seen when the electromagnetic interaction is considered that it also contributes to quark confinement and also to the strong force, so the coupling constant of the hadronic interaction is much less than the Yukawa value, and it is possible that perturbation theory can be used.

There appears to be scope for testing the model in the analysis of jets and in the predictions of the gyromagnetic moment of the proton and the neutron. Jets are are predicted to be formed out the the creation of hadron-antihadron pairs, not out of individual quarks or gluons. It is thought that jets will show some characteristic from the original particle production, and that such characteristics can in principle be analysed. On the grounds of a general similarity between this model and QCD we anticipate a value for the gyromagnetic moment of the proton and the neutron which is near to, but not the same as, the prediction of QCD. But, in common with QCD, to test it a very substantial calculation must be carried out without error. The calculation has not been attempted at the present time.

We will use the following notations defined in [1] and [2]. $S$ is the set of spin indices. $\mathcal{F}$ is the set of physically realisable states. The coordinate system is $N = (-v, v] \otimes (-v, v] \otimes (-v, v] \subset \mathbb{N}^3$ for some $v \in \mathbb{N}$, where $(-v, v] = \{x \in \mathbb{N} | -v < x \leq v\}$. A field operator is a mapping $\mathbb{R}^4 \otimes S \to \mathcal{F}$, where the elements of $\mathcal{F}$ are regarded as operators. The field operator for the creation of a particle in interaction is

1.1        $\forall (x, \alpha) = (x_0, \boldsymbol{x}, \alpha) \in \mathbb{R}^4 \otimes S \, . \, \underline{|x, \alpha\rangle} : \mathcal{F} \to \mathcal{F}$

The field operators for the of creation of the antiparticle is $\overline{|x, \alpha\rangle} : \mathcal{F} \to \mathcal{F}$. The field operators for the annihilation of particles in interaction are $\underline{\langle x, \alpha|} : \mathcal{F} \to \mathcal{F}$ and $\langle \overline{x, \alpha|} : \mathcal{F} \to \mathcal{F}$ and are the hermitian conjugates of the creation field operators.

## 2  Hadron Fields

It is postulated that all interactions can be represented as a number of particles annihilated or created at a point. Dirac particles are fermions, obeying the Pauli exclusion principle so an interaction cannot create an even number of the same Dirac particle. But there is nothing to stop the creation of a particle-antiparticle pair, or the creation of a triplet of three Dirac particles (provided that there is no two particle subsystem).

**Definition:** The baryon field operator creating three quarks, indexed 1,2,3, at $x \in \mathbb{R}^4$ is

2.1 $$| x, \underline{\alpha}_1, \underline{\alpha}_2, \underline{\alpha}_3 \rangle \equiv \underline{| x, \alpha_1, \alpha_2, \alpha_3 \rangle} \equiv \underline{| x, \alpha_1 \rangle}\,\underline{| x, \alpha_2 \rangle}\,\underline{| x, \alpha_3 \rangle}$$

**Definition:** The baryon field operator creating three antiquarks, indexed 1,2,3, at $x \in \mathbb{R}^4$ is

2.2 $$| x, \overline{\alpha}_1, \overline{\alpha}_2, \overline{\alpha}_3 \rangle \equiv \overline{| x, \alpha_1, \alpha_2, \alpha_3 \rangle} \equiv \overline{| x, \alpha_1 \rangle}\,\overline{| x, \alpha_2 \rangle}\,\overline{| x, \alpha_3 \rangle}$$

**Definition:** The meson field operator creating a quark, 1 and antiquark, 2 is

2.3 $$| x, \underline{\alpha}_1, \overline{\alpha}_2 \rangle \equiv \underline{| x, \alpha_1, \alpha_2 \rangle} \equiv \underline{| x, \alpha_1 \rangle}\,\overline{| x, \alpha_2 \rangle}$$

Interacting field operators make it explicit that in interaction the annihilation of a particle is equivalent to the creation of the corresponding antiparticle. They are regularised so that a quark cannot be created and annihilated in the same interaction.

**Definition:** For a baryon the interacting field operator is

2.4 $$\phi_{\alpha_1 \alpha_2 \alpha_3}(x) = | x, \overline{\alpha}_1, \overline{\alpha}_2, \overline{\alpha}_3 \rangle + \langle x, \underline{\alpha}_1, \underline{\alpha}_2, \underline{\alpha}_3 |$$

**Corollary:** The hermitian conjugate of the interacting baryon field is the antibaryon field

2.5 $$\phi^{\dagger}{}_{\alpha_1 \alpha_2 \alpha_3}(x) = \phi_{\overline{\alpha}_1 \overline{\alpha}_2 \overline{\alpha}_3}(x)$$

**Proof:** immediate from 2.4.

**Definition:** For a meson the interacting field operator is

2.6 $$\phi_{\alpha_1 \overline{\alpha}_2}(x) = | x, \overline{\alpha}_1, \underline{\alpha}_2 \rangle + \langle x, \underline{\alpha}_1, \overline{\alpha}_2 |$$

**Corollary:** The hermitian conjugate of the interacting meson field is also a meson field

2.7 $$\phi^{\dagger}{}_{\alpha_1 \overline{\alpha}_2}(x) = \phi_{\overline{\alpha}_1 \alpha_2}(x) = \phi_{\alpha_2 \overline{\alpha}_1}(x)$$

**Proof:** immediate from 2.6  and 2.3.

It is possible to postulate purely hadronic interactions from these field operators, corresponding to the Yukawa potential [8] in which a baryon absorbs or emits a meson (figure 2.9). Then the general form of the interaction operator is

2.8 $$\sum_{x \,\in\, \mathbb{N}} \phi_{\overline{\alpha}_1 \overline{\alpha}_2 \overline{\alpha}_3}(x) \phi_{\beta_3 \overline{\alpha}_4}(x) \phi_{\beta_1 \beta_2 \beta_4}(x) \Gamma_{\alpha_1 \alpha_2 \alpha_3 \alpha_4 \beta_1 \beta_2 \beta_3 \beta_4}$$

plus terms for the interaction with quarks in the first and second index positions. The summation convention is assumed for spin indices, and $\Gamma$ is some combination of Dirac $\gamma$ matrices. All possible quark combinations are included, subject to the condition that quark number is conserved in the interaction for each flavour of quark.

**Figure 2.9:** *Feynman diagram for a possible interaction in which a baryon absorbs or emits a meson. The (optional) heavy joining line indicates the particles meet at a point.*

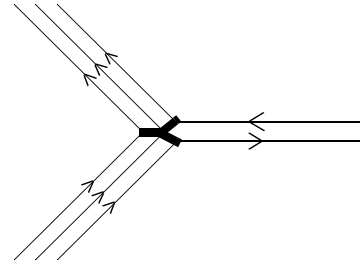

The corresponding meson interaction (figure 2.11) is

2.10 $$\sum_{x\,\in\,\mathbb{N}} \phi_{\overline{\alpha}_1\alpha_2}(x)\phi_{\overline{\beta}_2\alpha_3}(x)\phi_{\beta_1\overline{\beta}_3}(x)\Gamma_{\alpha_1\alpha_2\alpha_3\beta_1\beta_2\beta_3}$$

**Figure 2.11:** *Feynman diagram for a possible interaction in which a meson absorbs or emits a meson.*

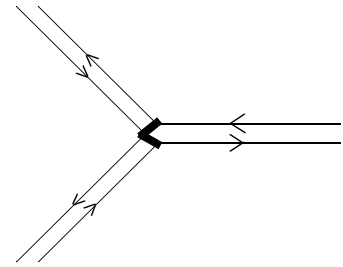

# 3 The Electromagnetic Interaction

The treatment of the electromagnetic interaction follows that for leptons, but there are now three Dirac adjoints for baryons, and two for mesons.

**Definition:** The Dirac adjoints of the baryon annihilation operator $\langle\, x\,,\underline{\alpha}_1,\,\underline{\alpha}_2,\,\underline{\alpha}_3\,|$ are

3.1 $$|\, x\,,\hat{\underline{\alpha}}_1,\,\underline{\alpha}_2,\,\underline{\alpha}_3\,\rangle \;=\; |\, x\,,\underline{\mu},\,\underline{\alpha}_2,\,\underline{\alpha}_3\rangle\gamma^0_{\mu\alpha_1} \qquad \text{(summation convention is used for } \mu)$$

3.2 $$|\, x\,,\underline{\alpha}_1,\,\hat{\underline{\alpha}}_2,\,\underline{\alpha}_3\,\rangle \;=\; |\, x\,,\underline{\alpha}_1,\,\underline{\mu},\,\underline{\alpha}_3\rangle\gamma^0_{\mu\alpha_2}$$

3.3 $$|\, x\,,\underline{\alpha}_1,\,\underline{\alpha}_2,\,\hat{\underline{\alpha}}_3\,\rangle \;=\; |\, x\,,\underline{\alpha}_1,\,\underline{\alpha}_2,\,\underline{\mu}\rangle\gamma^0_{\mu\alpha_3}$$

**Definition:** The Dirac adjoints of the meson annihilation operator $\langle\, x\,,\underline{\alpha}_1,\overline{\alpha}_2\,|$ are

3.4 $$|\, x\,,\underline{\alpha}_1,\hat{\overline{\alpha}}_2\,\rangle \;=\; |\, x\,,\underline{\alpha}_1,\overline{\mu}\rangle\gamma^0_{\mu\alpha_2}$$

3.5 $$|\, x\,,\hat{\underline{\alpha}}_1,\overline{\alpha}_2\,\rangle \;=\; |\, x\,,\underline{\mu},\overline{\alpha}_2\rangle\gamma^0_{\mu\alpha_1}$$

**Definition:** The Dirac adjoints of the baryon field $\phi_{\alpha_1\alpha_2\alpha_3}(x)$, 2.4, are

3.6 $$\phi_{\hat{\overline{\alpha}}_1\alpha_2\overline{\alpha}_3}(x) \;=\; \phi^\dagger_{\mu\alpha_2\alpha_3}(x)\gamma^0_{\mu\alpha_1} \;=\; |\, x\,,\hat{\underline{\alpha}}_1,\underline{\alpha}_2,\underline{\alpha}_3\,\rangle + \langle\, x\,,\hat{\overline{\alpha}}_1,\overline{\alpha}_2,\overline{\alpha}_3\,|$$

and similarly for the other two quarks $\phi_{\overline{\alpha}_1\hat{\alpha}_2\overline{\alpha}_3}(x)$ and $\phi_{\overline{\alpha}_1\hat{\alpha}_2\overline{\alpha}_3}(x)$.

**Definition:** The Dirac adjoints of the meson field $\phi_{\alpha_1\overline{\alpha}_2}(x)$, 2.6, are

3.7 $$\phi_{\hat{\overline{\alpha}}_1\alpha_2}(x) \;=\; (|\, x\,,\underline{\mu},\overline{\alpha}_2\rangle + \langle\, x\,,\overline{\mu},\underline{\alpha}_2|)\gamma^0_{\mu\alpha_1}$$

3.8 $$\phi_{\overline{\alpha}_1\hat{\alpha}_2}(x) \;=\; (|\, x\,,\underline{\alpha}_1,\overline{\mu}\rangle + \langle\, x\,,\overline{\alpha}_1,\underline{\mu}|)\gamma^0_{\mu\alpha_1}$$

The coupling of photon to quark in an electromagnetic interaction follows the pattern of the coupling of photon to lepton in QED. The photon only couples to one quark at a time, but all three quarks of a baryon, or both quarks of a meson are part of the field. Let the quarks indexed 1,2,3 have charges $e_1$, $e_2$, $e_3$. Then the electromagnetic interaction for a baryon (figure 3.10) is

3.9
$$\sum_{x \,\in\, N} :e_1 \phi_{\hat{\overline{\alpha}}_1 \overline{\alpha}_2 \overline{\alpha}_3}(x) \gamma_{\alpha_1 \alpha'_1} \cdot A(x) \phi_{\alpha'_1 \alpha_2 \alpha_3}(x)$$
$$+ e_2 \phi_{\overline{\alpha}_1 \hat{\overline{\alpha}}_2 \overline{\alpha}_3}(x) \gamma_{\alpha_2 \alpha'_2} \cdot A(x) \phi_{\alpha_1 \alpha'_2 \alpha_3}(x)$$
$$+ e_3 \phi_{\overline{\alpha}_1 \overline{\alpha}_2 \hat{\overline{\alpha}}_3}(x) \gamma_{\alpha_3 \alpha'_3} \cdot A(x) \phi_{\alpha_1 \alpha_2 \alpha'_3}(x):$$

where the colons reorder the expression by placing all creation operators to the left of all annihilation operators, to ensure that false values are not generated corresponding to the annihilation of particles in the interaction in which they are created.

**Figure 3.10:** *Feynman diagram for an interaction in which a baryon absorbs or emits a photon. The photon couples to one quark, but all three quarks must be present.*

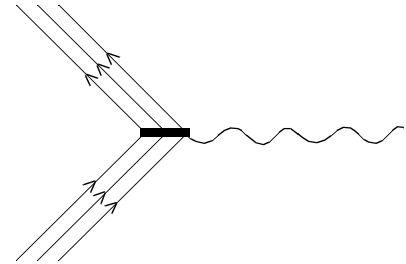

Similarly for a meson the electromagnetic interaction (figure 3.12) is

3.11
$$\sum_{x \,\in\, N} :e_1 \phi_{\hat{\overline{\alpha}}_1 \alpha_2}(x) \gamma_{\alpha_1 \alpha'_1} \cdot A(x) \phi_{\alpha'_1 \overline{\alpha}_2}(x)$$
$$- e_2 \phi_{\overline{\alpha}_1 \hat{\alpha}_2}(x) \gamma_{\alpha_2 \alpha'_2} \cdot A(x) \phi_{\alpha_1 \overline{\alpha}'_2}(x):$$

**Figure 3.12:** *Feynman diagram for an interaction in which a meson absorbs or emits a photon. The photon couples to one quark, but both quarks must be present.*

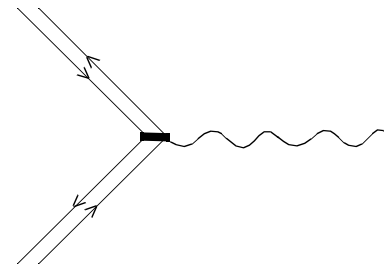

**Theorem:** In the electromagnetic interaction, in the case when the hadron is preserved (i.e. ignoring pair creation and pair annihilation) the photon interacts with one quark, and leaves spin and momentum of the other quark(s) unchanged.

**Proof:** It is sufficient to demonstrate the theorem for the first term of 3.9. The proof for the other terms of 3.9 and 3.11 is identical. The first term of 3.9 is

3.13
$$\sum_{x \,\in\, N} e_1 \phi_{\hat{\overline{\alpha}}_1 \overline{\alpha}_2 \overline{\alpha}_3}(x) \gamma_{\alpha_1 \alpha'_1} \cdot A(x) \phi_{\alpha'_1 \alpha_2 \alpha_3}(x)$$

Then using 3.6 and 2.4, 3.13 can be written

3.14
$$\sum_{x \in N} : e_1(|x, \hat{\underline{\alpha}}_1, \underline{\alpha}_2, \underline{\alpha}_3\rangle + \langle x, \hat{\bar{\alpha}}_1, \bar{\alpha}_2, \bar{\alpha}_3|)\gamma_{\alpha_1\alpha'_1} \cdot A(x)$$
$$(|x, \bar{\alpha}'_1, \bar{\alpha}_2, \bar{\alpha}_3\rangle + \langle x, \underline{\alpha}'_1, \underline{\alpha}_2, \underline{\alpha}_3|):$$

The term affecting a particle is

3.15
$$\sum_{x \in N} e_1|x, \hat{\underline{\alpha}}_1, \underline{\alpha}_2, \underline{\alpha}_3\rangle\gamma_{\alpha_1\alpha'_1} \cdot A(x)\langle x, \underline{\alpha}'_1, \underline{\alpha}_2, \underline{\alpha}_3|$$

which is, by 2.1, in matrix notation ([1] section 6, *Multiparticle States*).

3.16
$$\begin{bmatrix} \sum_{x \in N} e_1|x, \hat{\underline{\alpha}}_1\rangle\gamma_{\alpha_1\alpha'_1} \cdot A(x)\langle x, \underline{\alpha}'_1| \\ \sum_{x \in N} |x, \underline{\alpha}_2\rangle\langle x, \underline{\alpha}_2| \\ \sum_{x \in N} |x, \underline{\alpha}_3\rangle\langle x, \underline{\alpha}_3| \end{bmatrix}$$

In 3.16 it is seen that the first quark participates in an interaction with an identical form to the electromagnetic interaction for leptons, while the other two particles are unaffected, by the resolution of unity. An identical argument applies to the antiparticle term in 3.14, and to each of the terms in 3.9 and 3.11.

## 4 Consequences

### 4.1 The Strong Field

Just as the classical electromagnetic field consists of a cloud of "virtual" photons surrounding a charged particles, in the present model the strong field consists of "virtual" hadrons (principally mesons) surrounding a hadron. Virtual particles are perhaps more accurately thought of as real. The absence of knowledge about any individual virtual particles implies that they have a probabilistic effect on the behaviour of matter, which can be described in a field theoretic approach. The distinction between this and QCD is that here the strong field consists solely of quarks and anti-quarks created in interactions of the form of 2.8, 2.10, 3.9, and 3.11. We are not aware of any direct experimental test which can be carried out for the absence of gluons.

### 4.2 Isospin and SU(N)

The properties of SU(N) follow directly and immediately from the definition of multiparticle space given in [1], and the interactions 2.8 and 2.10 under the constraint on that possible quark combinations are included subject to the condition that quark number is conserved in the interaction for each flavour of quark. SU(N) is equivalent to the statement that in the absence of knowledge of the flavour of an individual quark, it is labelled by a ket formed from a linear combination of the possible flavours.

### 4.3 Zweig's Rule

It has been observed in experiment that, when a meson consists of a quark and its corresponding antiquark, the two do not immediately self-destruct. This follows directly from the interaction 2.10. As a result the mesons consisting of $s\bar{s}$, $c\bar{c}$, and $b\bar{b}$ combinations are relatively stable. Conservation of energy prevents their decay into other particles in which the quark and antiquark survive separately

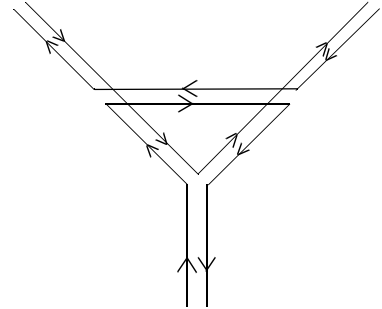

**Figure 4.4:** *Zweig's rule is not absolute, because the quark-antiquark pair can annihilate in a second order process. This requires greater energy than was available in the original meson, but takes place because conservation of energy is temporarily suspended.*

### 4.5 Superstrong Force

Quarks are confined in hadrons because there is only one space co-ordinate in the interaction density for all the quarks in an interaction 2.8, 2.10, 3.9 and 3.11. It is convenient to think that they share the same wave function in co-ordinate space; if one quark is localised, for example by an electromagnetic interaction with other matter, then the other quarks are confined at the same point when the interaction takes place. These interactions are taking place all the time so quarks are confined within baryons and mesons. Free quarks, if they exist, cannot participate in electromagnetic interactions, and therefore cannot be directly detected, but if they do exist we might expect certain interactions to take place more easily, and this may be testible.

### 4.6 Quasi-Free Quarks

The interaction generates a momentum conserving delta function for each individual quark. Thus for hadrons in which the quarks are in eigenstates of momentum, the interaction leaves the momenta and the spin states of two quarks in a baryon unchanged, as though only one of the three quarks participates in the interaction. Thus quarks are `quasi free' – they are confined in co-ordinate space, and yet have values of momentum independent of each other.

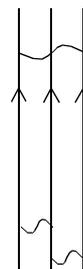

**Figure 4.7:** *Photons passing between the quarks in a baryon transfer momentum from one quark to another.*

The self interaction in which a photon is exchanged between two of the quarks in the baryon ensures that momentum can be transferred from one quark to another (figure 4.7). Quasi free quarks can also be understood by application of the uncertainty principle. If the quarks in a hadron are confined in coordinate space, their momenta are indeterminate, and hence independent. The prediction that quarks are quasi-free was originally observed experimentally, but in the absence of any other satisfactory model of quark confinement, we believe that it strongly supports the current model.

### 4.8   Quark Sharing

Only quark triplets and quark-antiquark pairs can interact, so only these combinations can be observed. But this does not resolve whether quarks are independent particles or whether they are parts of more complex particles with an indivisible structure. The resolution of this question depends on whether, once a quark triplet has been created, the same three quarks must come together to be annihilated in the next interaction, or whether any three quarks would do. We will now examine the qualitive properties of hadrons on the assumption that interactions can consist of any three quarks.

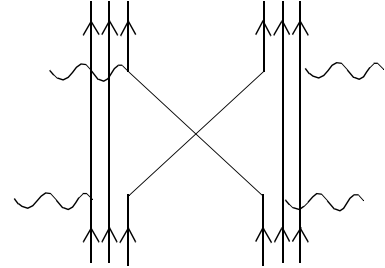

**Figure 4.9:** *If quarks are independent particles then it need not be the same quark triplets which participate in all interactions in the nucleus.*

The result of allowing any triplet of free quarks to combine in an interaction for a baryon is that quarks will be shared between baryons in the nucleus of the atom. Effectively this results in the transmission of mesons between hadrons, and conveys a Yukawa-type potential by means of electromagnetic interactions, considerably amplifying the strong force, or rather reducing the value of the coupling constant necessary to maintain it. Because the quarks in a hadron are closely bound it is expected that large numbers of photons are transmitted between them, and it is not immediately obvious how to calculate cross-sections or to predict magnitude of the strong binding force. In addition to electromagnetic interactions, the calculation of the strong force may involve purely hadronic interactions, as well as "identity interactions"

4.10 $$\sum_{x \,\in\, \mathrm{N}} \phi_{\overline{\alpha}_1 \overline{\alpha}_2 \overline{\alpha}_3}(x) \phi_{\alpha_1 \alpha_2 \alpha_3}(x) \ \text{ and } \ \sum_{x \,\in\, \mathrm{N}} \phi_{\overline{\alpha}_1 \alpha_2}(x) \phi_{\alpha_1 \overline{\alpha}_2}(x)$$

which enable quark sharing without the need for a photon. As with the second an third entries of 3.16, identity interactions leave the momenta and spin states of the individual quarks unchanged, while still requiring them to be confined in space.

### 4.11  Jets in Hadron Scattering

If hadrons are fired at each other at very high energies, they may exchange a meson (figure 4.12). The quarks within the resulting pair of hadrons have excessive energy differences and do not have a stable shared waved function. Thus the resulting hadrons rapidly decay. Because each of the two hadrons has an overall momentum along the axis the original collision, they decay into jets of hadrons moving on that axis. The energy and spread of the jets should be calculable from a knowledge of the (stable) state of quarks in the initial hadrons.

**Figure 4.12:** *Meson exchange by the exchange of two quarks..*

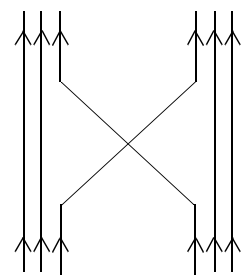

### 4.13 Jets in Electron Positron Scattering

A beam of positrons is fired at a beam of electrons at very high energy. A positron combines with an electron to create a single photon. The photon has zero momentum and seriously violates conservation of energy, so it must decay rapidly into other particles.

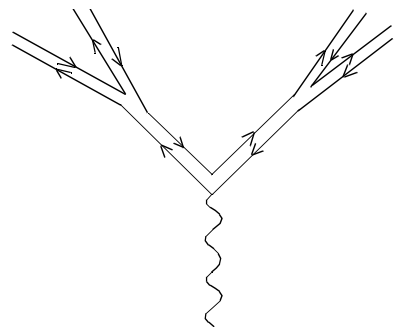

**Figure 4.14:** *An energy conservation violating photon creates a hadron antihadron pair. The hadrons have equal and opposite total momenta, and high internal energy, so they rapidly decay into jets of particles in opposite directions.*

This explanation of jets appears to us more successful than the usual interpretation in quantum chromodynamics, where it is suggested that jets arise from a single quark, and which fails to satisfactorily resolve the problem that if the jet arose from a single quark there should be a loose quark left over after the formation of hadrons. We expect jets not from single quarks, but as a result of any interaction which seriously violates energy conservation, as the interaction initially creates particles with high momenta to restore energy conservation as quickly as possible.

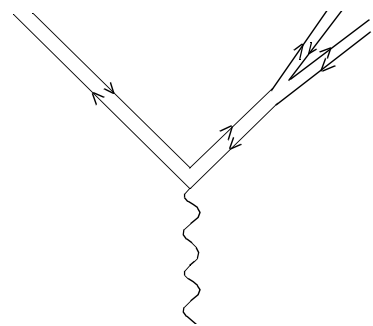

**Figure 4.15:** *According to the model, at high energies greater internal energy is dissipated in the first decay after the initial pair creation. I.e. this interaction does more to restore conservation of energy than subsequent ones. As energies increase, the residual inbalance in energy conservation first leads to a thickening of one of the jets, and then to a separate jet.*

If this model of jets is correct, then we expect to categorise patterns of jets according to the initial pair creation. A frequent decay would derive from a mesons consisting of two quarks of approximately equal mass. In this case two of the jets would derive from one meson and be approximately equal, and the other would tend to be of nearly double the energy. A less frequent occurence would be jets arising from the creation of an initial baryon-antibaryon pair, where the third jet will corresponding to a meson splitting of a baryon, and will generate a weaker jet. At any given energy, the frequency of each pattern of jets should correspond to the predicted frequency of each hadron pair creation, and can be calculated by standard methods.

### 4.16 Perturbation Theory.

In [1] we calculated Feynman rules and showed that the current discrete model of qed necessitates that we subtract a term the propagator which recognises that a particle cannot be annihilated at the instant of its creation. It was shown that when this term is subtracted finite results are a obtained in agreement with the standard predictions of qed after renormalisation. It is necessary to renormalise the electron mass to take into account the self interaction, but no infinite quantities are involved. The perturbation theoretic approach based on Feynman diagrams used in this paper gives qualitive agreement with experimental results. The mechanisms proposed here for quark binding and quark sharing predict a coupling constant with a value much less than that anticipated by the Yukawa potential based on simple meson exchange e.g. [9], so a perturbation theory approach to calculation may be technically feasible. But, although the metaphysic is much simpler than that of qcd or string theory there are still large numbers of diagrams and quantitive predictions are not so simple. In other respects the model shares predictions with standard theories based on general considerations such as analyticy, relativistic invariance, and symmetries, as well as the successful phenomenology other parton models [5].